\documentclass[showpacs,aps,graphicx,twocolumn]{revtex4}
\usepackage{amsmath}
\usepackage{graphicx}

\begin{document}

\title{Generation and complete nondestructive analysis of hyperentanglement assisted by nitrogen-vacancy centers in resonators }

\author{Qian Liu and Mei Zhang\footnote{Corresponding author: zhangmei@bnu.edu.cn} }

\address{Department of Physics, Applied Optics Beijing Area Major Laboratory,
Beijing Normal University, Beijing 100875, China}

\date{\today }

\begin{abstract}
We present two efficient schemes for the deterministic  generation
and the complete nondestructive analysis of hyperentangled Bell
states in both the polarization and spatial-mode degrees of freedom
(DOFs) of two-photon systems, assisted by  the nitrogen-vacancy (NV)
centers in diamonds coupled to microtoroidal resonators as a result
of cavity quantum electrodynamics (QED). With the input-output
process of photons, two-photon polarization-spatial hyperentangled
Bell states can be generated in a deterministic way and their
complete nondestructive analysis can be achieved. These schemes can
be generalized to generate and analyze hyperentangled
Greenberger-Horne-Zeilinger states of multi-photon systems as well.
Compared with previous works, these two schemes relax the difficulty
of their implementation in experiment as it is not difficult to
obtain the $\pi$ phase shift in single-sided NV-cavity systems.
Moreover, our schemes do not require that the transmission for the
uncoupled cavity is balanceable with the reflectance for the coupled
cavity. Our calculations show that these schemes can reach a high
fidelity and efficiency with current technology, which may be a
benefit to long-distance high-capacity quantum communication with
two DOFs of photon systems.
\end{abstract}

\maketitle



\section{Introduction}
\label{sec1}

Recently, hyperentanglement, which is defined as the entanglement in
multiple degrees of freedom (DOFs) of a quantum system
\cite{heper1,heper2,heper3}, has attracted much attention as it has
some important applications in quantum information processing. It
can speedup quantum computation (e.g., hyper-parallel photonic
quantum computation \cite{hypercomputation,hypercomputation2}) and
it can be used to assist the complete Bell-state analysis and
entanglement purification. In 2003, Walborn \emph{et al.}
\cite{Walborn} presented a simple linear-optical scheme for the
complete Bell-state analysis of photons with hyperentanglement. In
2006, Schuck \emph{et al.} \cite{Schuck} demonstrated the  complete
deterministic analysis of polarization Bell states with only linear
optics assisted by polarization-time-bin hyperentanglement. In 2007,
Barbieri \emph{et al.} \cite{Barbieri} demonstrated  the complete
and deterministic discrimination of polarization Bell states
assisted by momentum entanglement. In 2009, Wilde and Uskov
\cite{quantumec}  proposed a quantum-error-correcting-code scheme
assisted by linear-optical hyperentanglement.  In 2010, some
deterministic entanglement purification protocols (EPPs)
\cite{EPPsheng2,EPPsheng3,EPPdeng} were proposed with the
polarization-spatial hyperentanglement of photon systems, which can
solve the troublesome problem that  a large amount of quantum
resources would be sacrificed in the conventional EPPs, and they are
very useful in practical quantum repeaters. In 2012, Wang, Song, and
Long \cite{repeater4} proposed an important quantum repeater
protocol with polarization-spatial hyperentanglement.





The hyperentanglement of photon systems can also be used to largely
increase the channel capacity of quantum communication. For example,
Barreiro \emph{et al.} \cite{HESC} beat the channel capacity limit
of photonic superdense coding with
polarization-orbital-angular-momentum hyperentanglement in linear
optics in 2008.  In 2010, Sheng, Deng, and Long \cite{kerr} gave the
first  scheme for the complete hyperentangled-Bell-state analysis
(HBSA) for quantum communication, and they designed the pioneering
model for the quantum teleportation with two DOFs of photon pairs,
resorting to cross-Kerr nonlinearity. In 2011, Pisenti \emph{et al}.
\cite{HBSA1} pointed out the limitations for manipulation and
measurement of entangled systems with inherently linear unentangling
devices. In 2012, Ren \emph{et al.} \cite{HBSA2} proposed another
interesting scheme for the complete HBSA for photon systems by using
the giant nonlinear optics in quantum dot-cavity systems and they
presented the entanglement swapping with photonic
polarization-spatial hyperentanglement. In 2012, Wang, Lu, and Long
\cite{HBSA3} introduced an interesting scheme for the complete HBSA
for photon systems by the giant circular birefringence induced by
double-sided quantum-dot-cavity systems.  In 2013, Ren, Du, and Deng
\cite{HECP} proposed the parameter-splitting method to extract the
polarization-spatial maximally hyerentangled photons  when the
coefficients of the initial partially hyperentangled states are
known, and this fascinating method  can be achieved with the maximum
success probability by performing the protocol only once, resorting
to linear-optical elements only. They \cite{HECP} also gave the
first hyperentanglement concentration protocol (hyper-ECP) for
unknown polarization-spatial less-hyperentangled states with
linear-optical elements only \cite{HECP}. Ren and Deng
\cite{HyperEPP}  presented the first hyperentanglement purification
protocol (hyper-EPP) for two-photon systems in polarization-spatial
hyperentangled states, and it is very useful in the high-capacity
quantum repeaters with hyperentanglement.  In 2014, Ren, Du, and
Deng \cite{RenHEPP2} gave a two-step hyper-EPP for
polarization-spatial hyperentangled states with the
quantum-state-joining method, and it has a far higher efficiency. In
the same time,  Ren and Long \cite{hyperecpgeneral} proposed a
general hyper-ECP for photon systems assisted by quantum dot spins
inside optical microcavities. Recently, Li and Ghose  presented an
interesting hyper-ECP resorting to linear optics
\cite{lixhyperconcentration} and another efficient hyper-ECP for the
multipartite hyperentangled state via the cross-Kerr nonlinearity
\cite{lixhyperconcentration2}.


%

Another attractive candidate for solid-state quantum information
processing is the nitrogen-vacancy (NV) center in a diamond, owing
to its long decoherence time even at room temperature
\cite{Jelezko}, and its spin can be   initialized and readout via a
highly stable optical transition \cite{Davies,Gruber}. By using the
coherent manipulation of an electron spin and nearby individual
nuclear spins, Dutt \emph{et al.} \cite{register} demonstrated a
controllable quantum register in NV centers in 2007.
Decoherence-protected quantum gates for a hybrid solid-state
register \cite{register1} was also experimentally demonstrated on a
single NV center. As this system allows for high-fidelity
polarization and detection of single electron and nuclear spin
states even under ambient conditions \cite{PD1,PD2,Gruber,PD3}, the
multipartite entanglement among single spins in diamond was
demonstrated by Neumann \emph{et al.} \cite{Neumann} in   2008.  In
2010, Togan \emph{et al.} \cite{QEG} realized the quantum
entanglement generation of an optical photon and an NV center.
Photon Fock states on-demand can be implemented in a low-temperature
solid-state quantum system with an NV center in a nano-diamond
coupled to a nearby high-Q optical cavity \cite{Fock}.

%

%

Recently, a combination of NV centers and microcavities,  a
promising solid-state cavity quantum electrodynamics (QED) system,
has gained widespread attention
\cite{transfer,s33,s34,s35,s36,s37,s38,WeigateNV,wang}. One of the
microcavities is called microtoroidal resonator (MTR) with a
quantized whispering-gallery mode (WGM) and required to be of a high
Q factor and a small mode volume \cite{MTR1,MTR2}. However, when MTR
couples to a fiber, its Q factor is surely degraded \cite{s34}. The
single-photon input-output process from a MTR in experiment also has
been demonstrated \cite{s34}. In 2009, the quantum nondemolition
measurement on a single spin of an NV center has been proposed with
a low error rate \cite{s37} and it was experimentally demonstrated
through Faraday rotation \cite{s38} in 2010. In 2011, Chen \emph{et
al.} \cite{s36} proposed an efficient scheme to entangle separate NV
centers by coupling to MTRs. In 2013, Wei and Deng \cite{WeigateNV}
proposed some interesting schemes for compact quantum gates on
electron-spin qubits assisted by diamond NV centers inside cavities.


%


In this paper, we present two efficient schemes to generate
deterministically hyperentangled states, i.e., hyperentangled Bell
states and hyperentangled Greenberger-Horne-Zeilinger (GHZ) states,
in which photons are entangled in both the polarization and
spatial-mode DOFs, assisted by the NV centers in diamonds coupled to
MTRs  and  the input-output process of photons as a result of cavity
QED. We also propose a scheme to distinguish completely the 16
polarization-spatial hyperentangled Bell states, and it works in a
nondestructive way. After analyzing the hyperentangled Bell states,
the photon systems can be used for other tasks in quantum
information processing. Compared with previous works
\cite{kerr,HBSA2,HBSA3}, these two schemes relax the difficulty of
their implementation in experiment as it is not difficult to obtain
the $\pi$ phase shift in single-sided NV-cavity systems. Moreover,
they do not require that the transmission for the uncoupled cavity
is balanceable with the reflectance for the coupled cavity. Our
calculations show that these schemes can work with a high fidelity
and efficiency with current experimental techniques, which may be
beneficial to long-distance high-capacity quantum communication,
such as quantum teleportation, quantum dense coding, and  quantum
superdense coding with  two DOFs of photon systems.

This article is organized as follows. In Sec. \ref{sec2}, we will
introduce the diamond-NV-center system and its single-photon
input-output process. The generation of hyperentangled Bell states
and hyperentangled GHZ states, and the complete nondestructive HBSA
are presented in Secs. \ref{sec3} and \ref{sec4}. A discussion and a
summary are given in Sec. \ref{sec5}.

\begin{figure}[!h]
\begin{center}
\includegraphics[width=7.2 cm,angle=0]{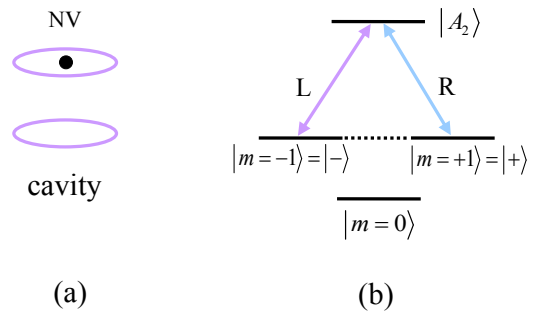}
\caption{(Color online) (a) Schematic diagram for an NV center
inside an MTR. (b) The electron energy-level configuration of an NV
center in an MTR with the relevant transitions driven by different
polarized photons. $L$ ($R$) represents the left(right) circularly
polarized photon. }\label{fig1}
\end{center}
\end{figure}

\section{A nitrogen-vacancy center in microtoroidal resonator}
\label{sec2}

As shown in Fig. \ref{fig1}(a), an NV center, composed of a
substitutional nitrogen atom and an adjacent vacancy in diamond
lattice, is coupled to an MTR  with a WGM. The NV center is
negatively charged with two unpaired electrons located at the
vacancy, and the energy-level structure of the NV center coupling to
the cavity mode is shown in Fig. \ref{fig1}(b). The ground state is
a spin triplet with the splitting at $2.87$ GHz between the levels
$|m=0\rangle$ and $|m=\pm1\rangle$ owing to spin-spin interactions.
The specifically excited state, which is one of the six eigenstates
of the full Hamitonian including spin-orbit and spin-spin
interactions in the absence of any perturbation, such as by an
external magnetic field or crystal strain, is labeled as \cite{QEG}
$|A_{2}\rangle=|E_{-}\rangle|+\rangle+|E_{+}\rangle|-\rangle$, where
$|E_{+}\rangle$ and $|E_{-}\rangle$ are the orbital states with the
angular momentum projections $+1$ and $-1$ along the NV axis,
respectively. The optical  transition is allowed between the ground
state $|m=\pm1\rangle$ and the excited state $|A_{2}\rangle$ owing
to the total angular momentum conservation \cite{QEG,Lenef}.

An NV center can be  modeled as a $\Lambda$-type three-level
structure with the ground state $|-\rangle=|m=-1\rangle$ and
$|+\rangle=|m=1\rangle$, and the excited state is
$|1\rangle=|A_{2}\rangle$. The transitions
$|-\rangle\leftrightarrow|1\rangle$ and
$|+\rangle\leftrightarrow|1\rangle$ in the NV center are resonantly
coupled to the right (R) and the left (L) circularly polarized
photons with the identical transition frequency, respectively.
Considering the structure in Fig.\ref{fig1}, which can be modeled as
a single-sided cavity, one can write down the Heisenberg equations
of motion for this system as follows:
\begin{eqnarray}    
\begin{split}
\frac{da}{dt}&=-[i(\omega_{c}-\omega)+\frac{\kappa}{2}+\frac{\kappa_{s}}{2}]\,a-\text{g}\sigma_{\!-}-\sqrt{\kappa}\,a_{in},\\
\frac{d\sigma_{\!-}}{dt}&=-[i(\omega_{0}-\omega)+\frac{\gamma}{2}]\sigma_{\!-}-\text{g}\sigma_{\!z}a,\\
a_{out}&=a_{in}+\sqrt{\kappa}\,a,
\end{split}
\end{eqnarray}
where $\omega$, $\omega_{c}$, and $\omega_{0}$ are the frequencies
of the photon, cavity mode, and the atomic-level transition,
respectively. $\text{g}$ is the coupling strength between the NV
center and the cavity mode. $\frac{\gamma}{2}$ and
$\frac{\kappa}{2}$ are the decay rates of the NV center and the
cavity field, respectively. $\frac{\kappa_{s}}{2}$ is the side
leakage rate of the cavity. $a_{in}$ and $a_{out}$ are the input and
the output field operators.

In the weak excitation approximation, the reflection coefficient in
the steady state can be obtained,
\begin{eqnarray}\label{eq1}    
r(\omega)=1-\frac{\kappa[i(\omega_{0}-\omega)+\frac{\gamma}{2}]}{[i(\omega_{0}-\omega)+
\frac{\gamma}{2}][i(\omega_{c}-\omega)+\frac{\kappa}{2}+\frac{\kappa_{s}}{2}]+\text{g}^{2}}.\nonumber\\
\end{eqnarray}
For $\text{g}=0$, the reflection coefficient $r_{0}(\omega)$ is
\begin{eqnarray}\label{eq2} 
r_{0}(\omega)=\frac{i(\omega_{c}-\omega)-\frac{\kappa}{2}+\frac{\kappa_{s}}{2}}{i(\omega_{c}-\omega)+\frac{\kappa}{2}+\frac{\kappa_{s}}{2}}.
\end{eqnarray}
From Eqs. (\ref{eq1}) and   (\ref{eq2}), one can see that if
$\omega_{0}=\omega_{c}=\omega$,
\begin{eqnarray}    
r(\omega)=\frac{(\kappa_{s}-\kappa)\gamma+4\text{g}^{2}}{(\kappa_{s}+\kappa)\gamma+4\text{g}^{2}},\;\;\;\;\;\;\;\;
r_{0}(\omega)=\frac{\kappa_{s}-\kappa}{\kappa_{s}+\kappa}.
\end{eqnarray}
If the NV center is in the initial state $|-\rangle$ ($|+\rangle$)
and a single polarized photon $|L\rangle$ ($|R\rangle$) is input,
the photon will experience a phase shift $e^{i\phi}$ owing to the
Faraday rotation. However, if the initial state of the NV center is
$|-\rangle$ ($|+\rangle$), the input photon with $|R\rangle$
($|L\rangle$) polarization will get a phase shift $e^{i\phi_{0}}$.
In the resonant condition $\omega_{0}=\omega_{c}=\omega$, when
$\kappa_{s}\ll\kappa$ and $4\text{g}^{2}\gg\kappa\gamma$, we
approximately have $\phi=0$ and $\phi_{0}=\pi$ from Eq. (\ref{eq1}).
The  change of the input photon  can be summarized as follows
\cite{WeigateNV}:
\begin{eqnarray}   
|R\rangle|+\rangle&\rightarrow&|R\rangle|+\rangle,\;\;\;\;\;\;\,
|R\rangle|-\rangle\;\rightarrow\;-|R\rangle|-\rangle,\nonumber\\
|L\rangle|+\rangle&\rightarrow&-|L\rangle|+\rangle,\;\;\;\;
|L\rangle|-\rangle\;\rightarrow\;|L\rangle|-\rangle.\label{eq5}
\end{eqnarray}

\section{Photonic hyperentanglement generation} \label{sec3}

We first describe how to generate two-photon polarization-spatial
hyperentangled Bell states assisted by NV centers coupled to MTRs as
a result of cavity QED, and then extend this approach for the
generation of three-photon polarization-spatial hyperentangled GHZ states.\\

\subsection{Generation of two-photon hyperentangled Bell states}
\label{sec31}

A two-photon hyperentangled Bell state in both the polarization and
the spatial-mode DOFs can be expressed as
\begin{eqnarray}   
|\eta_{1}^{+}\rangle_{PS}  =  \frac{1}{2}(|RR\rangle+|LL\rangle)_{ab} (|a_{1}b_{1}\rangle+|a_{2}b_{2}\rangle)_{ab}.
\end{eqnarray}
Here, $|R\rangle$ and $|L\rangle$ denote the right-circular
polarization and the left-circular polarization of photons,
respectively. $a_{1}$ ($b_{1}$) and $a_{2}$ ($b_{2}$) are the
different spatial modes for  photon $a$ (b). The subscripts $P$ and
$S$ denote the polarization and the spatial-mode DOFs, respectively.
$a$ and $b$ represent the two photons in the hyperentangled state.
The four Bell states in the polarization DOF can be expressed as
\begin{eqnarray}   
\begin{split}
|\Phi_{1}^{\pm}\rangle_{P}&=\frac{1}{\sqrt{2}}(|RR\rangle\pm|LL\rangle),\\
|\Phi_{2}^{\pm}\rangle_{P}&=\frac{1}{\sqrt{2}}(|LR\rangle\pm|RL\rangle),
\end{split}
\end{eqnarray}
and those in the spatial-mode DOF can be written as
\begin{eqnarray}  
\begin{split}
|\Phi_{1}^{\pm}\rangle_{S}&=\frac{1}{\sqrt{2}}(|a_{1}b_{1}\rangle\pm|a_{2}b_{2}\rangle),\\
|\Phi_{2}^{\pm}\rangle_{S}&=\frac{1}{\sqrt{2}}(|a_{2}b_{1}\rangle\pm|a_{1}b_{2}\rangle).
\end{split}
\end{eqnarray}

\begin{figure*}[tpb]  
\begin{center}
\includegraphics[width=13.6 cm,angle=0]{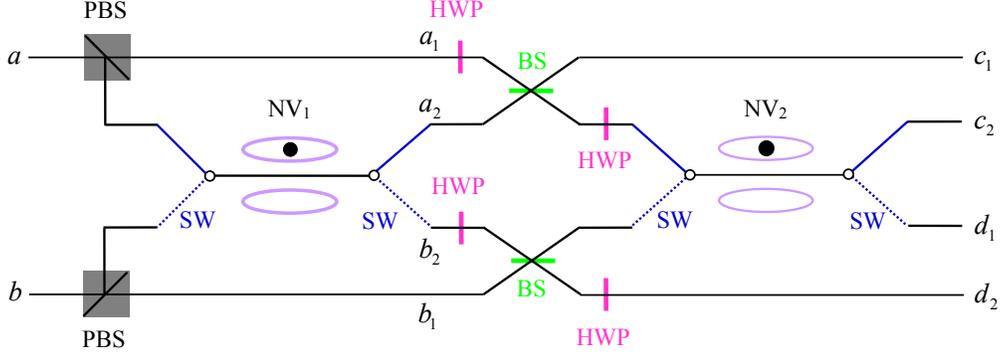}
\caption{(Color online) Schematic diagram for two-photon
polarization-spatial HBSG. PBS represents  a circular polarization
beam splitter which is used to transmit the $L$ polarized photon and
reflect  the $R$ polarized photon, respectively. SW represents an
optical switch and BS is a 50:50 beam splitter. HWP represents a
half-wave plate  which is used to perform a bit-flip operation
$X=|R\rangle\langle L|+|L\rangle\langle R|$ on the photon in the
polarization DOF. NV$_1$ and NV$_2$ represent two NV centers.  $k_1$
and $k_2$ ($k=a,b,c,d$) represent different spatial
modes.}\label{fig2}
\end{center}
\end{figure*}

The principle of our scheme for the two-photon polarization-spatial
hyperentangled Bell states generation (HBSG) assisted by NV centers
coupled to MTRs as a result of cavity QED  is shown in
Fig.\ref{fig2}. Here SW is an optical switch and BS represents a
$50:50$ beam splitter which can accomplish the following
transformation in the spatial-mode DOF of the photons,
\begin{eqnarray}            
\begin{split}
K_{a_{1}(b_{2})}^{\dag}&\rightarrow \frac{1}{\sqrt{2}}(K_{c_{1}(d_{1})}^{\dag}+K_{c_{2}(d_{2})}^{\dag}),\\
K_{a_{2}(b_{1})}^{\dag}&\rightarrow
\frac{1}{\sqrt{2}}(K_{c_{1}(d_{1})}^{\dag}-K_{c_{2}(d_{2})}^{\dag}).
\end{split}
\end{eqnarray}

Suppose that two NV centers NV$_1$ and NV$_2$ are initialed to  the
superposition states
$|\varphi^{+}\rangle_{1}=|\varphi^{+}\rangle_{2}=\frac{1}{\sqrt{2}}(|-\rangle+|+\rangle)$,
and the two photons $a$ and $b$ with the same frequency are prepared
in the same initial state
$|\phi\rangle_{a}=|\phi\rangle_{b}=\frac{1}{\sqrt{2}}(|R\rangle+|L\rangle)$.
Photons $a$ and $b$ are successively sent into the device shown in
Fig.\ref{fig2}. A time interval $\Delta t$ exists between two
photons, and $\Delta t$ should be less than the spin coherence time
$T$. When photon $a$ passes through the cavity, the optical switch
(SW) is switched to await photon $b$. After passing through the two
NV centers, the two photons $a$ and $b$ can be entangled with the
electron spins in the NV centers in the two cavities. The
corresponding transformations on the states can be described as
follows:
\begin{eqnarray}\label{eq10}  
\!\!\vert
\Phi\rangle_0\!\!\!\!&=&\!\!|\phi\rangle_{a}|\phi\rangle_{b}|\varphi^{+}\rangle_{1}|\varphi^{+}\rangle_{2}\nonumber\\
\!\!\!\!&\xrightarrow{P\!B\!S\!s}&\!\!\!\frac{1}{2}(|L\rangle|a_{1}\!\rangle
\!+\! |R\rangle|a_{2}\!\rangle)(|L\rangle|b_{1}\!\rangle
\!+\! |R\rangle|b_{2}\!\rangle)|\varphi^{+}\!\rangle_{1}|\varphi^{+}\!\rangle_{2},\nonumber\\
&\!\!\!\! \xrightarrow{NV_1}&\!\!\! \frac{1}{2}
[(|LL\rangle|a_{1}b_{1}\rangle+|RR\rangle|a_{2}b_{2}\rangle)|\varphi^{+}\rangle_{1}\nonumber\\
&&
-(|LR\rangle|a_{1}b_{2}\rangle+|RL\rangle|a_{2}b_{1}\rangle)|\varphi^{-}\rangle_{1} ]|\varphi^{+}\rangle_{2},\nonumber\\
&\!\!\!\! \xrightarrow{B\!S}&\!\!\!\frac{1}{4}
[|\Phi_{1}^{+}\rangle_{P}|\Phi_{1}^{-}\rangle_{S}|\varphi^{+}\rangle_{1}
+|\Phi_{1}^{-}\rangle_{P}|\Phi_{2}^{-}\rangle_{S}|\varphi^{+}\rangle_{1}\nonumber\\
&&
-|\Phi_{2}^{+}\rangle_{P}|\Phi_{1}^{+}\rangle_{S}|\varphi^{-}\!\rangle_{1}
\!+\!|\Phi_{2}^{-}\rangle_{P}|\Phi_{2}^{+}\rangle_{S}|\varphi^{-}\!\rangle_{1}]|\varphi^{+}\!\rangle_{2},\nonumber\\
&\!\!\!\!
\xrightarrow{NV_2\;}&\!\!\!\frac{1}{4}(-|\Phi_{1}^{-}\rangle_{P}|\Phi_{1}^{-}\rangle_{S}|\varphi^{+}\rangle_{1}|\varphi^{-}\rangle_{2}
\nonumber\\
&&+\,|\Phi_{1}^{-}\rangle_{P}|\Phi_{2}^{-}\rangle_{S}|\varphi^{+}\rangle_{1}|\varphi^{+}\rangle_{2}\nonumber\\
&&
+\,|\Phi_{2}^{-}\rangle_{P}|\Phi_{1}^{-}\rangle_{S}|\varphi^{-}\rangle_{1}|\varphi^{-}\rangle_{2}
\nonumber\\
&&+\,|\Phi_{2}^{-}\rangle_{P}|\Phi_{2}^{-}\rangle_{S}|\varphi^{-}\rangle_{1}|\varphi^{+}\rangle_{2}).
\end{eqnarray}
Here
$|\varphi^{\pm}\rangle=\frac{1}{\sqrt{2}}(|-\rangle\pm|+\rangle)$.

From Eq. (\ref{eq10}), one can see that the injecting photons $a$
and $b$ pass through PBS in sequence. Each of them is split into two
wave-packets. The part of photon $a$ ($b$) in the state $|L\rangle$
is transmitted to path $a_{1}$ ($b_{1}$), while the part in the
state $|R\rangle$ is reflected by PBS to path $a_{2}$ ($b_{2}$) and
subsequently it interacts with the first NV center (named NV$_1$).
The wave-packets from the spatial modes $a_{1}$ and $a_{2}$ ($b_{1}$
and $b_{2}$) are mixed at the beam splitter (BS). HWP is used to
keep the polarization state of the photon unchanged. After BSs, the
states of the two-photon systems are divided into two groups,
according to the state of   NV$_2$ when
 it is measured  with the basis $\{\vert \varphi^+\rangle, \vert
\varphi^-\rangle\}$. So do the outcomes of the measurement on
NV$_1$. The relationship between the measurement outcomes of these
two NV centers and the polarization-spatial hyperentangled Bell
states of the two photons is shown in Table \ref{table1}.

\begin{table}[htb]
\centering \caption{The relation between the outcomes of the two NV
centers and the final polarization-spatial hyperentangled Bell
states.}
\begin{tabular}{ccc}
\hline\hline
   NV$_1$                 &     $\;\;\;\;\;\;\;\;$ NV$_2$ $\;\;\;\;\;\;\;\;$     &   hyperentangled Bell states       \\
   \hline
$|\varphi^{+}\rangle_1$   &       $|\varphi^{-}\rangle_2$                      &   $|\Phi_{1}^{-}\rangle_{P}|\Phi_{1}^{-}\rangle_{S}$ \\
$|\varphi^{+}\rangle_1$   &       $|\varphi^{+}\rangle_2$                      &   $|\Phi_{1}^{-}\rangle_{P}|\Phi_{2}^{-}\rangle_{S}$ \\
$|\varphi^{-}\rangle_1$   &       $|\varphi^{-}\rangle_2$                      &   $|\Phi_{2}^{-}\rangle_{P}|\Phi_{1}^{-}\rangle_{S}$ \\
$|\varphi^{-}\rangle_1$   &       $|\varphi^{+}\rangle_2$                      &   $|\Phi_{2}^{-}\rangle_{P}|\Phi_{2}^{-}\rangle_{S}$ \\
\hline\hline
\end{tabular}\label{table1}
\end{table}

From Table \ref{table1}, one can see that if   NV$_1$ is in the
state $|\varphi^{+}\rangle_{1}$ and  NV$_2$ is in the state
$|\varphi^{-}\rangle_{2}$,  the two-photon system $ab$ is in the
hyperentangled Bell state
$|\Phi_{1}^{-}\rangle_{P}|\Phi_{1}^{-}\rangle_{S}$. When the two NV
centers are in the states $|\varphi^{+}\rangle_{1}$ and
$|\varphi^{+}\rangle_{2}$, respectively, the two-photon system is in
the hyperentangled Bell state
$|\Phi_{1}^{-}\rangle_{P}|\Phi_{2}^{-}\rangle_{S}$. The
hyperentangled Bell states
$|\Phi_{2}^{-}\rangle_{P}|\Phi_{1}^{-}\rangle_{S}$ and
$|\Phi_{2}^{-}\rangle_{P}|\Phi_{2}^{-}\rangle_{S}$ also correspond
to different combinations of the  states of the two NV centers.
Therefore, one can generate the polarization-spatial hyperentangled
Bell states of the two-photon system by measuring the states of the
two NV centers. By applying a Hadamard operation on the
electron-spin state, its states
$\frac{1}{\sqrt{2}}(|-\rangle+|+\rangle)$ and
$\frac{1}{\sqrt{2}}(|-\rangle-|+\rangle)$ can be rotated to
$|+\rangle$ and $|-\rangle$, respectively. The measurement on spin
readout can be achieved with the resonant optical excitation
\cite{Gaebel}. Other polarization-spatial hyperentangled Bell states
can be obtained in a similar way or by resorting to the single-qubit
operations and the acquired hyperentangled Bell state.


\begin{figure*}[tpb]  
\begin{center}
\includegraphics[width=14cm,angle=0]{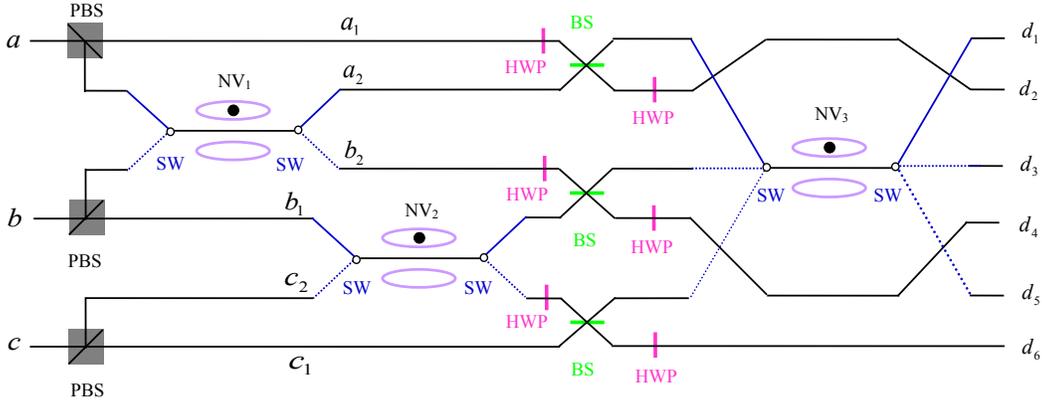}
\caption{(Color online) Schematic diagram  for the generation of
three-photon polarization-spatial  hyperentangled  GHZ
states.}\label{fig3}
\end{center}
\end{figure*}

\subsection{Generation of three-photon hyperentangled GHZ states}
\label{sec32}

Similar to the case for two-photon polarization-spatial
hyperentangled Bell states, we denote a three-photon
polarization-spatial hyperentangled GHZ state as
\begin{eqnarray}\label{eq11}
|\zeta_{1}^{+}\rangle_{PS} = \frac{1}{2}(|RRR\rangle\!+\!|LLL\rangle)_{abc}(|a_{1}b_{1}c_{1}\rangle\!+\!|a_{2}b_{2}c_{2}\rangle)_{abc}.\nonumber\\
\end{eqnarray}
Here $a$, $b$, and $c$ represent the three photons in the
hyperentangled state.  The  eight three-photon GHZ states in the
polarization DOF can be expressed as
\begin{eqnarray}\label{eq12}  
\begin{split}
|\Psi_{1}^{\pm}\rangle_{P}&=\frac{1}{\sqrt{2}}(|RRR\rangle\pm|LLL\rangle),
 \\
|\Psi_{2}^{\pm}\rangle_{P}&=\frac{1}{\sqrt{2}}(|LRR\rangle\pm|RLL\rangle), \\
|\Psi_{3}^{\pm}\rangle_{P}&=\frac{1}{\sqrt{2}}(|RLR\rangle\pm|LRL\rangle), \\
|\Psi_{4}^{\pm}\rangle_{P}&=\frac{1}{\sqrt{2}}(|RRL\rangle\pm|LLR\rangle).
\end{split}
\end{eqnarray}
and the eight GHZ states in the spatial-mode DOF are
\begin{eqnarray}\label{eq13} 
\begin{split}
|\Psi_{1}^{\pm}\rangle_{S}&=\frac{1}{\sqrt{2}}(|a_{1}b_{1}c_{1}\rangle\pm|a_{2}b_{2}c_{2}\rangle),\\
|\Psi_{2}^{\pm}\rangle_{S}&=\frac{1}{\sqrt{2}}(|a_{2}b_{1}c_{1}\rangle\pm|a_{1}b_{2}c_{2}\rangle),\\
|\Psi_{3}^{\pm}\rangle_{S}&=\frac{1}{\sqrt{2}}(|a_{1}b_{2}c_{1}\rangle\pm|a_{2}b_{1}c_{2}\rangle),\\
|\Psi_{4}^{\pm}\rangle_{S}&=\frac{1}{\sqrt{2}}(|a_{1}b_{1}c_{2}\rangle\pm|a_{2}b_{2}c_{1}\rangle).
\end{split}
\end{eqnarray}

The principle of our scheme for the generation of a three-photon
polarization-spatial  hyperentangled GHZ state is shown in
Fig.\ref{fig3}.  Considering all the three NV centers are in the
initial state
$|\varphi^{+}\rangle_{1}=|\varphi^{+}\rangle_{2}=|\varphi^{+}\rangle_{3}=\frac{1}{\sqrt{2}}(|-\rangle+|+\rangle)$,
and the flying photons $a$, $b$, and $c$ are in the superposition
state
$|\phi\rangle_{a}=|\phi\rangle_{b}=|\phi\rangle_{c}=\frac{1}{\sqrt{2}}(|R\rangle+|L\rangle)$.
A brief description of our scheme for hyperentangled-GHZ-state
generation can be written as follows.

Each of the three photons $a$, $b$, and $c$  is split into two
wave-packets by PBS. The photon in the state $|L\rangle$ is injected
into the pathes $a_{1}$, $b_{1}$, and $c_{1}$, while the photon in
the state $|R\rangle$ is sent into the pathes $a_{2}$,  $b_{2}$, and
$c_{2}$. The photons in the pathes $a_{2}$ and $b_{2}$ interact with
 NV$_1$, and those in the pathes $b_{1}$ and $c_{2}$ interact with
NV$_2$.   BSs mix the spatial modes $a_{1}$ and $a_{2}$, $b_{1}$ and
$b_{2}$, and $c_{1}$ and $c_{2}$. The states of the three photons
are divided into four groups, according to the states of NV$_1$ and
NV$_2$.  Under the condition that   NV$_3$ is imported, the
hyperentangled states of the photons can be determined by measuring
the states of the NV centers. The evolution of the whole system can
be described as
\begin{eqnarray}\label{eq14}
\vert \Psi\rangle_0\!\!\!\!&=&\!\!|\phi\rangle_{a}|\phi\rangle_{b}|\phi\rangle_{c}|\varphi^{+}\rangle_{1}|\varphi^{+}\rangle_{2}|\varphi^{+}\rangle_{3}\nonumber\\
\!\!\!\!\!\!&\xrightarrow{U_T}&\!\!
\frac{1}{2\sqrt{2}}(|\Psi_{1}^{+}\rangle_{P}|\Psi_{2}^{+}\rangle_{S})|\varphi^{+}\rangle_{1}|\varphi^{-}\rangle_{2}|\varphi^{-}\rangle_{3}  \nonumber\\
&&  -|\Psi_{1}^{+}\rangle_{P}|\Psi_{2}^{-}\rangle_{S}|\varphi^{+}\rangle_{1}|\varphi^{-}\rangle_{2}|\varphi^{+}\rangle_{3}\nonumber\\
&&  +|\Psi_{4}^{+}\rangle_{P}|\Psi_{2}^{+}\rangle_{S}|\varphi^{+}\rangle_{1}|\varphi^{+}\rangle_{2}|\varphi^{-}\rangle_{3} \nonumber\\
&&   +|\Psi_{4}^{+}\rangle_{P}|\Psi_{2}^{-}\rangle_{S}|\varphi^{+}\rangle_{1}|\varphi^{+}\rangle_{2}|\varphi^{+}\rangle_{3}\nonumber\\
&&  +|\Psi_{3}^{+}\rangle_{P}|\Psi_{2}^{+}\rangle_{S}|\varphi^{-}\rangle_{1}|\varphi^{+}\rangle_{2}|\varphi^{-}\rangle_{3} \nonumber\\
&&  -|\Psi_{3}^{+}\rangle_{P}|\Psi_{2}^{-}\rangle_{S}|\varphi^{-}\rangle_{1}|\varphi^{+}\rangle_{2}|\varphi^{+}\rangle_{3}\nonumber\\
&&  +|\Psi_{2}^{+}\rangle_{P}|\Psi_{1}^{+}\rangle_{S}|\varphi^{-}\rangle_{1}|\varphi^{-}\rangle_{2}|\varphi^{-}\rangle_{3} \nonumber\\
&&  -|\Psi_{2}^{+}\rangle_{P}|\Psi_{2}^{-}\rangle_{S}|\varphi^{-}\rangle_{1}|\varphi^{-}\rangle_{2}|\varphi^{+}\rangle_{3}).
\end{eqnarray}
Here $U_T$ represents the total operation by PBS, NV$_1$, NV$_2$,
BS, and NV$_3$ in sequence.
$|\varphi^{\pm}\rangle=\frac{1}{\sqrt{2}}(|-\rangle\pm|+\rangle)$.
The relation between the outcomes of the measurements on the three
NV centers and the obtained final polarization-spatial
hyperentangled GHZ states is shown in Table \ref{table2}.

\begin{table}[htb]
\centering \caption{The relation between the outcomes of the three
NV centers and the  final hyperentangled GHZ states}
\begin{tabular}{cccc}
\hline\hline
   NV$_1$ & $\;\;\;\; $ NV$_2$ $\;\;\; $ & $\;\;\; $ NV$_3$ $\;\;\;\; $  &  hyperentangled GHZ states         \\
   \hline
$|\varphi^{+}\rangle_{1}$&$|\varphi^{-}\rangle_{2}$&$|\varphi^{-}\rangle_{3}$&$|\Psi_{1}^{+}\rangle_{P}|\Psi_{2}^{+}\rangle_{S}$ \\
$|\varphi^{+}\rangle_{1}$&$|\varphi^{-}\rangle_{2}$&$|\varphi^{+}\rangle_{3}$&$|\Psi_{1}^{+}\rangle_{P}|\Psi_{2}^{-}\rangle_{S}$ \\
$|\varphi^{+}\rangle_{1}$&$|\varphi^{+}\rangle_{2}$&$|\varphi^{-}\rangle_{3}$&$|\Psi_{4}^{+}\rangle_{P}|\Psi_{2}^{+}\rangle_{S}$ \\
$|\varphi^{+}\rangle_{1}$&$|\varphi^{+}\rangle_{2}$&$|\varphi^{+}\rangle_{3}$&$|\Psi_{4}^{+}\rangle_{P}|\Psi_{2}^{-}\rangle_{S}$ \\
$|\varphi^{-}\rangle_{1}$&$|\varphi^{+}\rangle_{2}$&$|\varphi^{-}\rangle_{3}$&$|\Psi_{3}^{+}\rangle_{P}|\Psi_{2}^{+}\rangle_{S}$ \\
$|\varphi^{-}\rangle_{1}$&$|\varphi^{+}\rangle_{2}$&$|\varphi^{+}\rangle_{3}$&$|\Psi_{3}^{+}\rangle_{P}|\Psi_{2}^{-}\rangle_{S}$ \\
$|\varphi^{-}\rangle_{1}$&$|\varphi^{-}\rangle_{2}$&$|\varphi^{-}\rangle_{3}$&$|\Psi_{2}^{+}\rangle_{P}|\Psi_{1}^{+}\rangle_{S}$ \\
$|\varphi^{-}\rangle_{1}$&$|\varphi^{-}\rangle_{2}$&$|\varphi^{+}\rangle_{3}$&$|\Psi_{2}^{+}\rangle_{P}|\Psi_{2}^{-}\rangle_{S}$ \\
\hline\hline
\end{tabular}\label{table2}
\end{table}

Table \ref{table2} shows that if the three NV centers are in the
states $|\varphi^{+}\rangle_{1}$, $|\varphi^{-}\rangle_{2}$, and
$|\varphi^{-}\rangle_{3}$, respectively, the three-photon system
$abc$ is in the polarization-spatial hyperentangled GHZ state
$|\Phi_{1}^{+}\rangle_{P}|\Phi_{2}^{+}\rangle_{S}$. When the NV
centers are in the state $|\varphi^{+}\rangle_{1}$,
$|\varphi^{-}\rangle_{2}$, and $|\varphi^{+}\rangle_{3}$, the final
hyperentangled GHZ state of the three photons is
$|\Phi_{1}^{+}\rangle_{P}|\Phi_{2}^{-}\rangle_{S}$. Our scheme can
in principle generate the eight deterministic hyperentangled
three-photon GHZ states by measuring the states of the NV centers.

\section{Complete  nondestructive polarization-spatial hyperentangled-Bell-state analysis}
\label{sec4}

\begin{figure*}[tpb]  
\begin{center}
\includegraphics[width=16cm,angle=0]{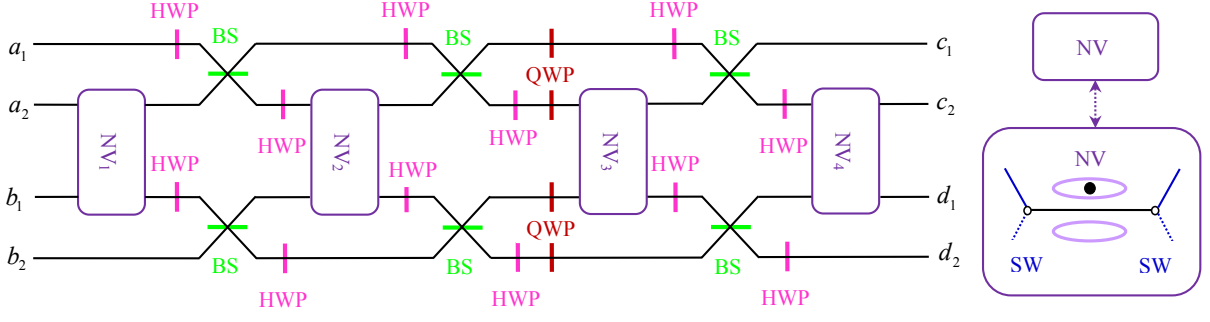}
\caption{(Color online) Schematic diagram for the complete
nondestructive analysis of two-photon polarization-spatial
hyperentangled
 Bell states. }\label{fig4}
\end{center}
\end{figure*}

The principle of our scheme for distinguishing the 16 hyperentangled
Bell states in both the polarization and the spatial-mode DOFs of
entangled photon pairs is shown  in Fig. \ref{fig4}. Here, we use
four NV centers coupled to four MTRs  and a few linear-optical
elements to achieve the complete  nondestructive analysis of
hyperentangled Bell states. Considering that two NV centers are both
initialized in the superposition state
$|\varphi^{+}\rangle_{1}=|\varphi^{+}\rangle_{2}=\frac{1}{\sqrt{2}}(|-\rangle+|+\rangle)$,
and a hyperentangled photon pair  is in one of the 16 hyperentangled
Bell states. As  shown in Fig. \ref{fig4}, one can let photon $a$
pass through the cavities first and then  photon $b$. SW is used for
the arrival of photon $b$ after photon $a$ passes through the
cavity. $\Delta t$ is the time interval between photon $a$ and
photon $b$ which is smaller than the spin coherence time of the NV
center.

According to  Eq.(\ref{eq5}), after the photons pass through NV$_1$
and NV$_2$, the evolution of the system composed of the two photons
and the two NV centers is shown in Table \ref{table3}. One can see
that the 16 polarization-spatial hyperentangled states can be
divided into four groups. If
 both the spins of  NV$_1$ and NV$_2$ are changed  to be
$|\varphi^{-}\rangle_{1}|\varphi^{-}\rangle_{2}$,  the photon pair
$ab$ is  in one of the four hyperentangled Bell states
$|\Phi_{1}^{\pm}\rangle_{P}|\Phi_{1}^{+}\rangle_{S}$ and
$|\Phi_{2}^{\pm}\rangle_{P}|\Phi_{1}^{-}\rangle_{S}$. If  both
NV$_1$ and NV$_2$  stay  in the initial state
$|\varphi^{+}\rangle_{1}|\varphi^{+}\rangle_{2}$, the photon pair
$ab$  is in one of the four hyperentangled Bell states
$|\Phi_{1}^{\pm}\rangle_{P}|\Phi_{2}^{-}\rangle_{S}$ and
$|\Phi_{2}^{\pm}\rangle_{P}|\Phi_{2}^{+}\rangle_{S}$. Otherwise,
when NV$_1$ and NV$_2$  are in the state of
$|\varphi^{-}\rangle_{1}|\varphi^{+}\rangle_{2}$ (or
$|\varphi^{+}\rangle_{1}|\varphi^{-}\rangle_{2}$), the photon pair
$ab$ is in one of the four hyperentangled Bell states
$|\Phi_{2}^{\pm}\rangle_{P}|\Phi_{1}^{+}\rangle_{S}$ and
$|\Phi_{1}^{\pm}\rangle_{P}|\Phi_{1}^{-}\rangle_{S}$ (or
$|\Phi_{1}^{\pm}\rangle_{P}|\Phi_{2}^{+}\rangle_{S}$ and
$|\Phi_{2}^{\pm}\rangle_{P}|\Phi_{2}^{-}\rangle_{S}$).

\begin{table}[htb]
\centering \caption{The relation between the initial states   and
the final states of the system after the photons pass through
NV$_{1}$ and NV$_{2}$.}
\begin{tabular}{cc}
\hline\hline
 $\;\;\; $   Initial states    $\;\;\; $  &      $\;\;\;\;\;\;\;\;\;$ Final states  $\;\;\; $        \\
   \hline
$|\Phi_{1}^{\pm}\rangle_{P}|\Phi_{1}^{+}\rangle_{S}|\varphi^{+}\rangle_{1}|\varphi^{+}\rangle_{2}$
&$\;\;\;\;$  $|\Phi_{1}^{\pm}\rangle_{P}|\Phi_{1}^{+}\rangle_{S}|\varphi^{-}\rangle_{1}|\varphi^{-}\rangle_{2}$ \\
$|\Phi_{2}^{\pm}\rangle_{P}|\Phi_{1}^{+}\rangle_{S}|\varphi^{+}\rangle_{1}|\varphi^{+}\rangle_{2}$
&$\;\;\;\;$  $|\Phi_{2}^{\mp}\rangle_{P}|\Phi_{2}^{-}\rangle_{S}|\varphi^{-}\rangle_{1}|\varphi^{+}\rangle_{2}$ \\
$|\Phi_{1}^{\pm}\rangle_{P}|\Phi_{1}^{-}\rangle_{S}|\varphi^{+}\rangle_{1}|\varphi^{+}\rangle_{2}$
&$\;\;\;\;$  $|\Phi_{1}^{\mp}\rangle_{P}|\Phi_{2}^{+}\rangle_{S}|\varphi^{-}\rangle_{1}|\varphi^{+}\rangle_{2}$ \\
$|\Phi_{2}^{\pm}\rangle_{P}|\Phi_{1}^{-}\rangle_{S}|\varphi^{+}\rangle_{1}|\varphi^{+}\rangle_{2}$
&$\;\;\;\;$  $|\Phi_{2}^{\pm}\rangle_{P}|\Phi_{1}^{-}\rangle_{S}|\varphi^{-}\rangle_{1}|\varphi^{-}\rangle_{2}$ \\
$|\Phi_{1}^{\pm}\rangle_{P}|\Phi_{2}^{+}\rangle_{S}|\varphi^{+}\rangle_{1}|\varphi^{+}\rangle_{2}$
&$\;\;\;\;$  $|\Phi_{1}^{\mp}\rangle_{P}|\Phi_{1}^{-}\rangle_{S}|\varphi^{+}\rangle_{1}|\varphi^{-}\rangle_{2}$ \\
$|\Phi_{2}^{\pm}\rangle_{P}|\Phi_{2}^{+}\rangle_{S}|\varphi^{+}\rangle_{1}|\varphi^{+}\rangle_{2}$
&$\;\;\;\;$  $|\Phi_{2}^{\pm}\rangle_{P}|\Phi_{2}^{+}\rangle_{S}|\varphi^{+}\rangle_{1}|\varphi^{+}\rangle_{2}$ \\
$|\Phi_{1}^{\pm}\rangle_{P}|\Phi_{2}^{-}\rangle_{S}|\varphi^{+}\rangle_{1}|\varphi^{+}\rangle_{2}$
&$\;\;\;\;$  $|\Phi_{1}^{\pm}\rangle_{P}|\Phi_{2}^{-}\rangle_{S}|\varphi^{+}\rangle_{1}|\varphi^{+}\rangle_{2}$ \\
$|\Phi_{2}^{\pm}\rangle_{P}|\Phi_{2}^{-}\rangle_{S}|\varphi^{+}\rangle_{1}|\varphi^{+}\rangle_{2}$
&$\;\;\;\;$  $|\Phi_{2}^{\mp}\rangle_{P}|\Phi_{1}^{+}\rangle_{S}|\varphi^{+}\rangle_{1}|\varphi^{-}\rangle_{2}$ \\
\hline\hline
\end{tabular}\label{table3}
\end{table}

Subsequently, one can let  the photons  pass through the BS and QWP
which are used to transform the phase information discrimination to
parity information discrimination both for the polarization DOF and
the spatial-mode DOF. According to Table \ref{table3}, after the
photons pass through  BS and  QWP, the initial hyperentangled Bell
state $|\Phi_{1}^{\pm}\rangle_{P}|\Phi_{1}^{+}\rangle_{S}$ and
$|\Phi_{2}^{\pm}\rangle_{P}|\Phi_{1}^{-}\rangle_{S}$ for the same
group will become
\begin{eqnarray}\label{eq16}
|\Phi_{1}^{+}\rangle_{P}|\Phi_{1}^{+}\rangle_{S}&\xrightarrow[]{NV_1,
NV_2, BS, QWP}&
|\Phi_{1}^{+}\rangle_{P}|\Phi_{1}^{+}\rangle_{S},\nonumber\\
|\Phi_{1}^{-}\rangle_{P}|\Phi_{1}^{+}\rangle_{S}&\xrightarrow[]{NV_1,
NV_2, BS, QWP}&
|\Phi_{2}^{+}\rangle_{P}|\Phi_{1}^{+}\rangle_{S},\nonumber\\
|\Phi_{2}^{+}\rangle_{P}|\Phi_{1}^{-}\rangle_{S}&\xrightarrow[]{NV_1,
NV_2, BS, QWP}&
|\Phi_{1}^{-}\rangle_{P}|\Phi_{2}^{+}\rangle_{S},\nonumber\\
|\Phi_{2}^{-}\rangle_{P}|\Phi_{1}^{-}\rangle_{S}&\xrightarrow[]{NV_1,
NV_2, BS, QWP}&
|\Phi_{2}^{-}\rangle_{P}|\Phi_{2}^{+}\rangle_{S}.\;\;\;\;
\end{eqnarray}
From Table \ref{table3}, one can see that the final hyperentangled
Bell states of two-photon systems in the same group will be changed
into four different groups.  By letting the photons pass through
NV$_3$ and NV$_4$ which has the same syndetic connection with NV$_1$
and NV$_2$, one can distinguish those hyperentangled Bell states.
The hyperentangled Bell states in the other three groups will have
the same conditions. After passing through all  the   elements shown
in Fig. \ref{fig4}, the final states of the two photons become those ones shown
in Table \ref{table4}.

\begin{table}[htb]
\centering \caption{The relation  between the initial hyperentangled
Bell states and the final states of the two photons after passing
through all  the  elements.}
\begin{tabular}{cccc}
\hline\hline
    Initial states    $\;  $  &       Final states   &  $\;  $  Initial states    $\;  $  &      $\; $ Final states        \\
   \hline
$|\Phi_{1}^{+}\rangle_{P}|\Phi_{1}^{+}\rangle_{S}$ &$\; $
$|\Phi_{1}^{+}\rangle_{P}|\Phi_{1}^{+}\rangle_{S}$ &$\; $
$|\Phi_{1}^{-}\rangle_{P}|\Phi_{1}^{+}\rangle_{S}$      &$\; $    $|\Phi_{2}^{-}\rangle_{P}|\Phi_{2}^{-}\rangle_{S}$ \\
$|\Phi_{2}^{+}\rangle_{P}|\Phi_{1}^{+}\rangle_{S}$      & $\; $
$|\Phi_{2}^{-}\rangle_{P}|\Phi_{1}^{+}\rangle_{S}$ &$\; $
$|\Phi_{2}^{-}\rangle_{P}|\Phi_{1}^{+}\rangle_{S}$      &$\; $    $|\Phi_{1}^{-}\rangle_{P}|\Phi_{2}^{-}\rangle_{S}$ \\
$|\Phi_{1}^{+}\rangle_{P}|\Phi_{1}^{-}\rangle_{S}$      &$\; $
$|\Phi_{2}^{+}\rangle_{P}|\Phi_{1}^{-}\rangle_{S}$ &$\; $
$|\Phi_{1}^{-}\rangle_{P}|\Phi_{1}^{-}\rangle_{S}$      &$\; $    $|\Phi_{1}^{-}\rangle_{P}|\Phi_{2}^{+}\rangle_{S}$ \\
$|\Phi_{2}^{+}\rangle_{P}|\Phi_{1}^{-}\rangle_{S}$      &$\; $
$|\Phi_{1}^{+}\rangle_{P}|\Phi_{1}^{-}\rangle_{S}$ &$\; $
$|\Phi_{2}^{-}\rangle_{P}|\Phi_{1}^{-}\rangle_{S}$      &$\; $    $|\Phi_{2}^{-}\rangle_{P}|\Phi_{2}^{+}\rangle_{S}$ \\
$|\Phi_{1}^{+}\rangle_{P}|\Phi_{2}^{+}\rangle_{S}$      &$\; $
$|\Phi_{2}^{+}\rangle_{P}|\Phi_{2}^{+}\rangle_{S}$ &$\; $
$|\Phi_{1}^{-}\rangle_{P}|\Phi_{2}^{+}\rangle_{S}$      &$\; $    $|\Phi_{1}^{-}\rangle_{P}|\Phi_{1}^{-}\rangle_{S}$ \\
$|\Phi_{2}^{+}\rangle_{P}|\Phi_{2}^{+}\rangle_{S}$      &$\; $
$|\Phi_{1}^{+}\rangle_{P}|\Phi_{2}^{+}\rangle_{S}$ &$\; $
$|\Phi_{2}^{-}\rangle_{P}|\Phi_{2}^{+}\rangle_{S}$      &$\; $    $|\Phi_{2}^{-}\rangle_{P}|\Phi_{1}^{-}\rangle_{S}$ \\
$|\Phi_{1}^{+}\rangle_{P}|\Phi_{2}^{-}\rangle_{S}$      &$\; $
$|\Phi_{1}^{+}\rangle_{P}|\Phi_{2}^{-}\rangle_{S}$ &$\; $
$|\Phi_{1}^{-}\rangle_{P}|\Phi_{2}^{-}\rangle_{S}$      &$\; $    $|\Phi_{2}^{-}\rangle_{P}|\Phi_{1}^{+}\rangle_{S}$ \\
$|\Phi_{2}^{+}\rangle_{P}|\Phi_{2}^{-}\rangle_{S}$      &$\; $
$|\Phi_{2}^{+}\rangle_{P}|\Phi_{2}^{-}\rangle_{S}$ &$\; $
$|\Phi_{2}^{-}\rangle_{P}|\Phi_{2}^{-}\rangle_{S}$      &$\; $    $|\Phi_{1}^{-}\rangle_{P}|\Phi_{1}^{+}\rangle_{S}$ \\
\hline\hline
\end{tabular}\label{table4}
\end{table}

In this time, by using linear-optical elements, one can transform
the  state of the two-photon system into its initial hyperentangled
Bell state. That is, one can completely distinguish the 16
hyperentangled Bell states with the measurement outcomes of the
states of the four NV centers rather than using single-photon
detectors to proceed destructive measurement. The relation
between the initial hyperentangled Bell states and the measurement
outcomes of the states of the NV centers is shown in Table
\ref{table5}.

\begin{table}[htb]
\centering \caption{The relation  between the initial hyperentangled
Bell states and the measurement outcomes of the states of the NV
centers.}
\begin{tabular}{ccccc}
\hline\hline
   Bell states         &     $\;\;$ NV$_1$ $\;\;$     &   $\;\;$  NV$_2$ $\;\;$    &   $\;\;$  NV$_3$ $\;\;$    &   $\;\;$   NV$_4$  $\;\;$         \\
   \hline
$|\Phi_{1}^{+}\rangle_{P}|\Phi_{1}^{+}\rangle_{S}$   &   $|\varphi^{-}\rangle_{1}$                  & $|\varphi^{-}\rangle_{2}$                & $|\varphi^{-}\rangle_{3}$                 & $|\varphi^{-}\rangle_{4}$ \\
$|\Phi_{1}^{-}\rangle_{P}|\Phi_{1}^{+}\rangle_{S}$   &   $|\varphi^{-}\rangle_{1}$                  & $|\varphi^{-}\rangle_{2}$                & $|\varphi^{-}\rangle_{3}$                 & $|\varphi^{+}\rangle_{4}$ \\
$|\Phi_{2}^{+}\rangle_{P}|\Phi_{1}^{+}\rangle_{S}$   &   $|\varphi^{-}\rangle_{1}$                  & $|\varphi^{+}\rangle_{2}$                & $|\varphi^{+}\rangle_{3}$                 & $|\varphi^{-}\rangle_{4}$ \\
$|\Phi_{2}^{-}\rangle_{P}|\Phi_{1}^{+}\rangle_{S}$   &   $|\varphi^{-}\rangle_{1}$                  & $|\varphi^{+}\rangle_{2}$                & $|\varphi^{+}\rangle_{3}$                 & $|\varphi^{+}\rangle_{4}$ \\
$|\Phi_{1}^{+}\rangle_{P}|\Phi_{1}^{-}\rangle_{S}$   &   $|\varphi^{-}\rangle_{1}$                  & $|\varphi^{+}\rangle_{2}$                & $|\varphi^{-}\rangle_{3}$                 & $|\varphi^{-}\rangle_{4}$ \\
$|\Phi_{1}^{-}\rangle_{P}|\Phi_{1}^{-}\rangle_{S}$   &   $|\varphi^{-}\rangle_{1}$                  & $|\varphi^{+}\rangle_{2}$                & $|\varphi^{-}\rangle_{3}$                 & $|\varphi^{+}\rangle_{4}$ \\
$|\Phi_{2}^{+}\rangle_{P}|\Phi_{1}^{-}\rangle_{S}$   &   $|\varphi^{-}\rangle_{1}$                  & $|\varphi^{-}\rangle_{2}$                & $|\varphi^{+}\rangle_{3}$                 & $|\varphi^{-}\rangle_{4}$ \\
$|\Phi_{2}^{-}\rangle_{P}|\Phi_{1}^{-}\rangle_{S}$   &   $|\varphi^{-}\rangle_{1}$                  & $|\varphi^{-}\rangle_{2}$                & $|\varphi^{+}\rangle_{3}$                 & $|\varphi^{+}\rangle_{4}$ \\
$|\Phi_{1}^{+}\rangle_{P}|\Phi_{2}^{+}\rangle_{S}$   &   $|\varphi^{+}\rangle_{1}$                  & $|\varphi^{-}\rangle_{2}$                & $|\varphi^{+}\rangle_{3}$                 & $|\varphi^{+}\rangle_{4}$ \\
$|\Phi_{1}^{-}\rangle_{P}|\Phi_{2}^{+}\rangle_{S}$   &   $|\varphi^{+}\rangle_{1}$                  & $|\varphi^{-}\rangle_{2}$                & $|\varphi^{+}\rangle_{3}$                 & $|\varphi^{-}\rangle_{4}$ \\
$|\Phi_{2}^{+}\rangle_{P}|\Phi_{2}^{+}\rangle_{S}$   &   $|\varphi^{+}\rangle_{1}$                  & $|\varphi^{+}\rangle_{2}$                & $|\varphi^{-}\rangle_{3}$                 & $|\varphi^{+}\rangle_{4}$ \\
$|\Phi_{2}^{-}\rangle_{P}|\Phi_{2}^{+}\rangle_{S}$   &   $|\varphi^{+}\rangle_{1}$                  & $|\varphi^{+}\rangle_{2}$                & $|\varphi^{-}\rangle_{3}$                 & $|\varphi^{-}\rangle_{4}$ \\
$|\Phi_{1}^{+}\rangle_{P}|\Phi_{2}^{-}\rangle_{S}$   &   $|\varphi^{+}\rangle_{1}$                  & $|\varphi^{+}\rangle_{2}$                & $|\varphi^{+}\rangle_{3}$                 & $|\varphi^{+}\rangle_{4}$ \\
$|\Phi_{1}^{-}\rangle_{P}|\Phi_{2}^{-}\rangle_{S}$   &   $|\varphi^{+}\rangle_{1}$                  & $|\varphi^{+}\rangle_{2}$                & $|\varphi^{+}\rangle_{3}$                 & $|\varphi^{-}\rangle_{4}$ \\
$|\Phi_{2}^{+}\rangle_{P}|\Phi_{2}^{-}\rangle_{S}$   &   $|\varphi^{+}\rangle_{1}$                  & $|\varphi^{-}\rangle_{2}$                & $|\varphi^{-}\rangle_{3}$                 & $|\varphi^{+}\rangle_{4}$ \\
$|\Phi_{2}^{-}\rangle_{P}|\Phi_{2}^{-}\rangle_{S}$   &   $|\varphi^{+}\rangle_{1}$                  & $|\varphi^{-}\rangle_{2}$                & $|\varphi^{-}\rangle_{3}$                 & $|\varphi^{-}\rangle_{4}$ \\
\hline\hline
\end{tabular}\label{table5}
\end{table}

From the analysis above, one can see that the hyperentangled Bell
states in both the polarization and the spatial-mode DOFs can be
completely distinguished assisted by NV centers in diamonds confined
in MTRs, and our analysis is nondestructive. Our scheme can be
generalized to the complete analysis of multi-photon
polarization-spatial  hyperentangled GHZ states by importing more
NV-center-cavity systems.

\section{Discussion and summary}\label{sec5}

\begin{figure}[!h]
\begin{center}
\includegraphics[width=7.5cm,angle=0]{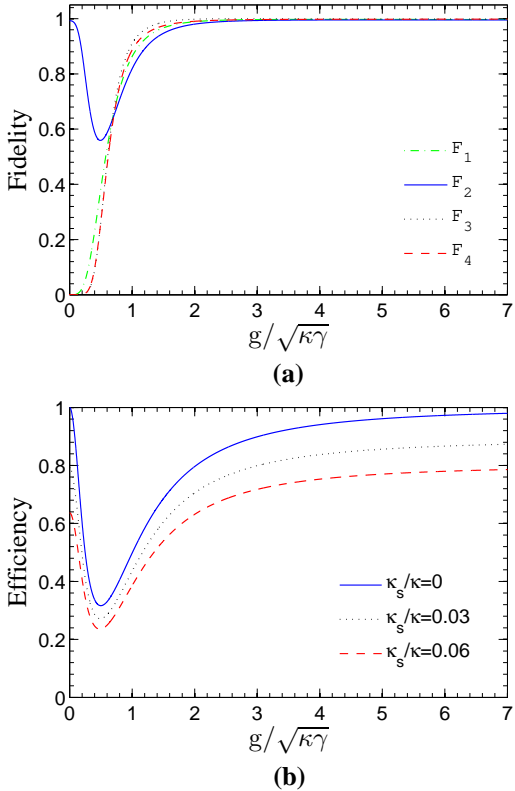}
\caption{(Color online) (a) The fidelities of the HBSG scheme for
the generation of the hyperentangled states
$|\Phi_{1}^{\pm}\rangle_{P}|\Phi_{1}^{\pm}\rangle_{S}$ ($F_{1}$),
$|\Phi_{1}^{\pm}\rangle_{P}|\Phi_{2}^{\pm}\rangle_{S}$ ($F_{2}$),
$|\Phi_{2}^{\pm}\rangle_{P}|\Phi_{1}^{\pm}\rangle_{S}$ ($F_{3}$),
and $|\Phi_{2}^{\pm}\rangle_{P}|\Phi_{2}^{\pm}\rangle_{S}$ ($F_{4}$)
vs the parameter $\text{g}/\sqrt{\kappa\gamma}$ for the leakage
rates $\kappa_{s}/\kappa=0.03$, respectively.  (b) The efficiency of
the HBSG scheme vs the parameter $\text{g}/\sqrt{\kappa\gamma}$ for
different leakage rates $\kappa_{s}/\kappa$.}\label{fig5}
\end{center}
\end{figure}

In our schemes, the reflection coefficient of input photon pulse and
the phase shift induced on the output photon play a crucial role.
Under the resonant condition $\omega_{0}=\omega_{c}=\omega$, if the
cavity side leakage is neglected, the fidelities of our schemes can
reach 100\% in the strong-coupling regime with $r(\omega)\cong1$ and
$r_{0}(\omega)\cong-1$. If the cavity leakage is taken into account,
the spin-selective optical transition rules employed in our work
become
\begin{eqnarray}   
|R\rangle|+\rangle&\rightarrow&r|R\rangle|+\rangle,\;\;\;\;\;\;\,
|R\rangle|-\rangle\;\rightarrow\;r_{0}|R\rangle|-\rangle,\nonumber\\
|L\rangle|+\rangle&\rightarrow&r_{0}|L\rangle|+\rangle,\;\;\;\;
|L\rangle|-\rangle\;\rightarrow\;r|L\rangle|-\rangle.\label{eq16}
\end{eqnarray}
Considering the practical implementation of the system,  we
numerically simulate the relation between the fidelities (the
efficiencies) and the coupling strength $\text{g}$, the cavity decay
rate $\kappa$, and the NV center decay rate $\gamma$. Defining the
fidelity of the process for generating or completely analyzing
hyperentangled states  in our schemes as
$F=|\langle\psi_{f}|\psi\rangle|^{2}$. Here $|\psi_{f}\rangle$ is
the final state by considering the cavity side leakage and
$|\psi\rangle$ denotes the final state with an ideal condition. We
calculate the fidelity of our scheme for generating the
hyperentangled state
$|\Phi_{1}^{-}\rangle_{P}|\Phi_{1}^{-}\rangle_{S}$ and its fidelity
is
\begin{eqnarray}\label{eq17}
F_{1}=\frac{(r-r_{0})^{2}(r^{2}+r_{0}^{2}+2)^{2}}{8(r^{2}+r_{0}^{2})(r^{4}+r_{0}^{4}+2)}.
\end{eqnarray}
By computations it is found, since the hyperentangled Bell states
$|\Phi_{1}^{+}\rangle_{P}|\Phi_{1}^{+}\rangle_{S}$,
$|\Phi_{1}^{+}\rangle_{P}|\Phi_{1}^{-}\rangle_{S}$, and
$|\Phi_{1}^{-}\rangle_{P}|\Phi_{1}^{+}\rangle_{S}$ have the same
parity conditions with the hyperentangled Bell state
$|\Phi_{1}^{-}\rangle_{P}|\Phi_{1}^{-}\rangle_{S}$, the  fidelities
of our scheme for their generations are also $F_{1}$. $F_{2}$,
$F_{3}$, and $F_{4}$ correspond to the fidelities for generating the
hyperentangled Bell states
$|\Phi_{1}^{\pm}\rangle_{P}|\Phi_{2}^{\pm}\rangle_{S}$,
$|\Phi_{2}^{\pm}\rangle_{P}|\Phi_{1}^{\pm}\rangle_{S}$, and
$|\Phi_{2}^{\pm}\rangle_{P}|\Phi_{2}^{\pm}\rangle_{S}$,
respectively. Here
\begin{eqnarray}\label{eq18}
\begin{split}
F_{2}&=\frac{(r^{2}+r_{0}^{2}+2)^{4}}{16(r^{4}+r_{0}^{4}+2)^{2}}, \\
F_{3}&=\frac{(r-r_{0})^{4}}{4(r^{2}+r_{0}^{2})^{2}}, \\
F_{4}&=\frac{(1-rr_{0})^{2}(r-r_{0})^{2}}{4(1+r^{2}r_{0}^{2})(r^{2}+r_{0}^{2})}.
\end{split}
\end{eqnarray}
The fidelities   of our HBSG scheme varies with  the parameter
$\text{g}/\sqrt{\kappa\gamma}$, shown in
 Fig. \ref{fig5} (a). For our HBSG scheme, the efficiency,
which is
 defined as the ratio of the number of the output photons to the
 input photons, can be written as
\begin{eqnarray}\label{eq19}
\eta_{1}=\frac{1}{2^{8}}(r^{2}+r_{0}^{2}+2)^{4}.
\end{eqnarray}
The efficiency of our HBSG scheme varies with  the parameter
$\text{g}/\sqrt{\kappa\gamma}$, shown in
 Fig. \ref{fig5} (b).

\begin{figure}[!h]
\begin{center}
\includegraphics[width=7.5cm,angle=0]{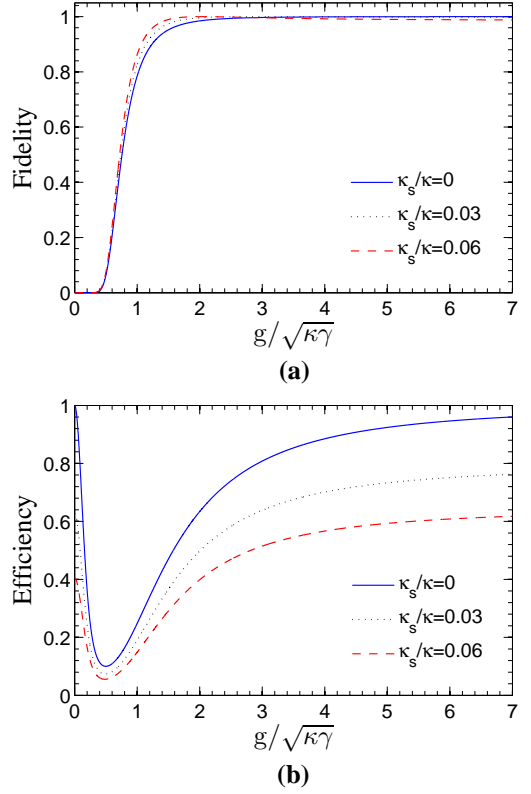}
\caption{(Color online) (a) Fidelity of the present HBSA scheme for
the hyperentangled Bell state
$|\Phi_{1}^{+}\rangle_{P}|\Phi_{1}^{+}\rangle_{S}$ versus the
parameter $\text{g}/\sqrt{\kappa\gamma}$ for different leakage rates
$\kappa_{s}/\kappa$. (b) The efficiency of the HBSA scheme vs the
parameter $\text{g}/\sqrt{\kappa\gamma}$ for different leakage rates
$\kappa_{s}/\kappa$.}\label{fig6}
\end{center}
\end{figure}

From Fig. \ref{fig5}(a), one can see that when
$\text{g}/\sqrt{\kappa\gamma}=1.5$, the fidelities are
$F_{1}=96.68\%$, $F_{2}=94.69\%$, $F_{3}=98.70\%$ , and
$F_{1}=97.43\%$ for the leakage rates $\kappa_{s}/\kappa=0.03$, and
the efficiency of our scheme is $\eta_{1}=53.96\%$ for the leakage
rates $\kappa_{s}/\kappa=0.06$. When the parameter
$\text{g}/\sqrt{\kappa\gamma}$ is larger than 3, the fidelities of
our HBSG scheme for the hyperentangled Bell states
$|\Phi_{2}^{\pm}\rangle_{P}|\Phi_{1}^{\pm}\rangle_{S}$ will be
$F_{3}\approx1$, while other  fidelities are  higher than $99\%$,
and the efficiency of our HBSG scheme will be higher than
$\eta_{1}=71.75\%$. In our protocol for generating hyperentangled
GHZ states, we use three NV-center-cavity systems rather than two
NV-center-cavity systems, which makes the fidelity and the
efficiency of the scheme lower than those in the HBSG protocol.

The fidelity of our HBSA protocol for the hyperentangled Bell state
$|\Phi_{1}^{+}\rangle_{P}|\Phi_{1}^{+}\rangle_{S}$ is given by
\begin{eqnarray}
F=\frac{\varepsilon^{8}}{4(\alpha+2\beta)}.
\end{eqnarray}
Here
\begin{eqnarray}
\begin{split}
\alpha &=(r^{2}-r_{0}^{2})^{2}[(r-r_{0})^{2}(r^{2}+1)(r_{0}^{2}+1)\\
&\;\;\;\; +4rr_{0}(r^{2}+r_{0}^{2})],\\
\beta  &=(r^{2}+r_{0}^{2})^{2}[(r^{2}+r_{0}^{2})^{2}+4r^{2}r_{0}^{2}],\\
\varepsilon &=r-r_{0}.
\end{split}
\end{eqnarray}
The efficiency of our HBSA protocol is
\begin{eqnarray}
\eta=\frac{1}{2^{16}}(r^{2}+r_{0}^{2}+2)^{8}.
\end{eqnarray}
Both the fidelity and efficiency of our HBSA scheme vary with the
parameter $\text{g}/\sqrt{\kappa\gamma}$, shown in Fig. \ref{fig6}.
The plots indicates that for $\text{g}/\sqrt{\kappa\gamma}\geq0.5$
the higher fidelities and efficiencies of our HBSG scheme depend on
the higher NV-cavity coupling strength. When the parameter
$\text{g}/\sqrt{\kappa\gamma}$ is larger than 3, the fidelity and
the efficiency will be higher than $F=99.58\%$ and $\eta=51.48\%$
for the leakage rates $\kappa_{s}/\kappa=0.06$.

The fidelity of our HBSG and HBSA schemes can be reduced a few
percent because of the imperfection of electron spin population, the
imperfection of the NV electron spins readout, and the imperfection
of frequency-selective microwave manipulation. For $m_{s}=0$, the
preparation fidelity is $99.7\pm0.1\%$, and $99.2\pm0.1\%$ for
$m_{s}=\pm1$ \cite{tgaebel}. The average readout fidelity of the NV
electron spins is $93.2\pm0.5\%$ \cite{tgaebel}. By using
isotopically purified diamonds and polarizing the nitrogen nuclear
spin, we can reduce the microwave manipulation imperfection. The
spin decoherence may also reduce the fidelity of our schemes. In our
schemes, the single photons are successively sent into the devices,
the time interval $\Delta t$ between two photons should be less than
the electron-spin coherence time $T$. In experimental cases, the
spin relaxation time $T_{1}$ of NV centers in diamond scales from
microseconds to seconds at low temperature and the dephasing time
$T_{2}$ is about 2 ms in an isotopically pure diamond
\cite{Neumann,r43}. The electron-spin coherence time $T$ is $>10$ ms
\cite{HBernien}, which is much longer than the photon coherence time
$\sim10$ ns and the subnanosecond electron-spin manipulation control
\cite{GDFuchs}.

For the practical  operations, the photon loss due to absorption and
scatting, the mismatch between the input pulse mode and MTR, and the
inefficiency of the detectors will bring ineffectiveness to our
schemes. Interestingly, the photon loss will just affect the success
efficiency, rather than the fidelity. The present experimental
technology can generate 300000 high-quality single photons within 30
s \cite{MH}. Another key ingredient of our protocols is the coupling
between NV centers and MTRs. In realistic experiments, the strong
coupling between the NV centers and the WGM has been demonstrated in
different kinds of microcavities
\cite{s33,nvcavity2,nvcavity3,nvcavity4,nvcavity5,nvcavity6}. The
coupling strength between NV centers and the WGM can reach
$\text{g}/2\pi\sim0.3-1$GHz
\cite{s33,nvcavity3,nvcavity4,nvcavity5}. The Q factor of the
chip-based microcavity is higher than 25000. Considering the
parameters $[\text{g}_{ZPL}, \kappa, \gamma_{total},
\gamma_{ZPL}]/2\pi=[0.30, 26, 0.013, 0.0004]$ GHz of an NV center
coupled to a microdisk \cite{nvcavity3}, we have
$\text{g}\approx3\sqrt{\kappa\gamma}$ and the fidelities of our HBSG
and HBSA schemes can exceed 99\%. Therefore our protocols are
feasible in experiment.

Compared with previous works \cite{kerr,HBSA2,HBSA3}, these two
schemes relax the difficulty of their implementation in experiment
as it is not difficult to generate the $\pi$ phase shift in
single-sided NV-cavity systems. Moreover, single-sided NV-cavity
systems have a long  coherence time even at the room temperature
(1.8 ms) \cite{r43}, different from quantum-dot-cavity systems. The
first HBSA scheme by Sheng, Deng, and Long \cite{kerr} is achieved
with cross-Kerr nonlinearity. It is perfect in theory.  At present,
a clean cross-Kerr nonlinearity in the optical single-photon regime
is still  a controversial assumption with current technology
\cite{kerr2,kerr3}. The second HBSA scheme by Ren \emph{et al.}
\cite{HBSA2} with single-sided quantum-dot-cavity systems  requires
the $\pi$ phase difference of the Faraday rotation between a hot
cavity and an empty cavity, and it is not easy to acquire the phase
difference  with only one nonlinear interaction between a photon and
a quantum dot. Compared with the work by Wang, Lu, and Long
\cite{HBSA3}, our schemes do not require that the transmission for
the uncoupled cavity is balanceable with the reflectance for the
coupled cavity. Moreover, the coherent manipulation of the spin of a
single NV center to accomplish quantum information and computation
tasks at room temperature has been presented
\cite{nvregister,algorithm}, which provides the basis for the
current schemes.

In summary, we have proposed two efficient schemes for the
deterministic generation and the complete nondestructive analysis of
hyperentangled Bell states in both polarization and spatial-mode
DOFs assisted by NV centers in MTRs. The HBSG protocol can also be
extended to achieve the generation of multi-photon hyperentangled
GHZ states efficiently. Compared with previous works
\cite{kerr,HBSA2,HBSA3}, our schemes relax the difficulty of their
implementation in experiment.  Our calculations show that the
proposed schemes can work with a high fidelity and efficiency under
the current experimental techniques, which may be a benefit to
long-distance high-capacity quantum communication, such as quantum
teleportation, quantum dense coding, and quantum superdense coding
with two DOFs of photon systems.

\section*{ACKNOWLEDGMENTS}

This work is supported by the National
Natural Science Foundation of China under Grant No. 11475021, the
National Key Basic Research Program of China under Grant No.
2013CB922000.

\end{document}